\documentclass[twocolumn]{aastex63}

\pdfoutput=1 
\usepackage{appendix}
\usepackage{amsmath,amstext}
\usepackage[T1]{fontenc}
\usepackage[figure,figure*]{hypcap}
\newcommand{\mos}{\,m\,s$^{-1}$}
\newcommand{\kms}{\,km\,s$^{-1}$}

\accepted{20 July 2021 in The Astronomical Journal}

\shorttitle{Spin-Orbit Alignment for AU\,Mic \MakeLowercase{b}}
\shortauthors{Addison et al.}

\graphicspath{{./}{figures/}}

\begin{document}

\title{AU\,Mic\,\MakeLowercase{b} is the Youngest Planet to have a Spin-Orbit Alignment Measurement}

\correspondingauthor{Brett C. Addison}
\email{Brett.Addison@usq.edu.au}

\author[0000-0003-3216-0626]{Brett C. Addison}
\affiliation{University of Southern Queensland, Centre for Astrophysics, USQ Toowoomba, West Street, QLD 4350 Australia}

\author[0000-0002-1160-7970]{Jonathan Horner}
\affiliation{University of Southern Queensland, Centre for Astrophysics, USQ Toowoomba, West Street, QLD 4350 Australia}

\author[0000-0001-9957-9304]{Robert A. Wittenmyer}
\affiliation{University of Southern Queensland, Centre for Astrophysics, USQ Toowoomba, West Street, QLD 4350 Australia}

\author[0000-0002-8091-7526]{Alexis Heitzmann}
\affiliation{University of Southern Queensland, Centre for Astrophysics, USQ Toowoomba, West Street, QLD 4350 Australia}

\author{Peter Plavchan}
\affiliation{Department of Physics \& Astronomy, George Mason University, 4400 University Drive MS 3F3, Fairfax, VA 22030, USA}

\author{Duncan J. Wright}
\affiliation{University of Southern Queensland, Centre for Astrophysics, USQ Toowoomba, West Street, QLD 4350 Australia}

\author[0000-0003-1360-4404]{Belinda A. Nicholson}
\affiliation{Sub-department of Astrophysics, Department of Physics, University of Oxford, Denys Wilkinson Building, Keble Road, Oxford, OX1 3RH, UK}
\affiliation{University of Southern Queensland, Centre for Astrophysics, USQ Toowoomba, West Street, QLD 4350 Australia}

\author[0000-0001-6208-1801]{Jonathan P. Marshall}
\affiliation{Academia Sinica Institute of Astronomy and Astrophysics, 11F of AS/NTU Astronomy-Mathematics Building,\\No.1, Sect. 4, Roosevelt Rd, Taipei 10617, Taiwan}
\affiliation{University of Southern Queensland, Centre for Astrophysics, USQ Toowoomba, West Street, QLD 4350 Australia}

\author[0000-0003-3964-4658]{Jake T. Clark}
\affiliation{University of Southern Queensland, Centre for Astrophysics, USQ Toowoomba, West Street, QLD 4350 Australia}

\author[0000-0002-3164-9086]{Maximilian N.\ G{\"u}nther}
\affiliation{Department of Physics and Kavli Institute for Astrophysics and Space Research, Massachusetts Institute of Technology, Cambridge, MA 02139, USA}
\affiliation{Juan Carlos Torres Fellow}

\author[0000-0002-7084-0529]{Stephen R. Kane}
\affiliation{Department of Earth and Planetary Sciences, University of California, Riverside, CA 92521, USA}

\author[0000-0003-3618-7535]{Teruyuki Hirano}
\affiliation{Department of Earth and Planetary Sciences, Tokyo Institute of Technology, 2-12-1 Ookayama, Meguro-ku, Tokyo 152-8551, Japan}

\author[0000-0002-7846-6981]{Songhu Wang}
\affiliation{Astronomy Department, Indiana University Bloomington, 727 East 3rd Street, Swain West 318, IN 4740, USA}

\author{John Kielkopf}
\affil{Department of Physics and Astronomy, University of Louisville, Louisville, KY 40292, USA}

\author[0000-0002-1836-3120]{Avi Shporer}
\affil{Department of Physics and Kavli Institute for Astrophysics and Space Research, Massachusetts Institute of Technology, Cambridge, MA 02139, USA}

\author{C.G. Tinney}
\affil{Exoplanetary Science at UNSW, School of Physics, UNSW Sydney, NSW 2052, Australia}

\author{Hui Zhang}
\affil{School of Astronomy and Space Science, Key Laboratory of Modern Astronomy and Astrophysics in Ministry of Education, Nanjing University, Nanjing 210046, Jiangsu, China}

\author{Sarah Ballard}
\affiliation{Department of Astronomy, University of Florida, 211 Bryant Space Science Center, Gainesville, FL, 32611, USA}


\author{Brendan P. Bowler}
\affil{Department of Astronomy, The University of Texas at Austin, TX 78712, USA}

\author[0000-0002-7830-6822]{Matthew W. Mengel}
\affil{University of Southern Queensland, Centre for Astrophysics, West Street, Toowoomba, QLD 4350 Australia}

\author{Jack Okumura}
\affil{University of Southern Queensland, Centre for Astrophysics, West Street, Toowoomba, QLD 4350 Australia}

\author[0000-0002-5258-6846]{Eric Gaidos}
\affiliation{Department of Earth Sciences, University of Hawai'i at M\={a}noa, Honolulu, Hawaii 96822 USA}

\author{Xian-Yu Wang}
\affiliation{National Astronomical Observatories, Chinese Academy of Sciences, Chaoyang District, Beijing, China}

\begin{abstract}
We report measurements of the sky-projected spin-orbit angle for AU\,Mic\,b, a Neptune-size planet orbiting a very young ($\sim20$\,Myr) nearby pre-main sequence M dwarf star which also hosts a bright, edge-on, debris disk. The planet was recently discovered from preliminary analysis of radial velocity observations and confirmed to be transiting its host star from photometric data from the NASA's \textit{TESS} mission. We obtained radial velocity measurements of AU\,Mic over the course of two partially observable transits and one full transit of planet b from high-resolution spectroscopic observations made with the {\textsc{Minerva}}-Australis telescope array. Only a marginal detection of the Rossiter--McLaughlin effect signal was obtained from the radial velocities, in part due to AU Mic being an extremely active star and the lack of full transit coverage plus sufficient out-of-transit baseline. As such, a precise determination of the obliquity for AU\,Mic\,b is not possible in this study and we find a sky-projected spin-orbit angle of $\lambda = 47{^{+26}_{-54}}^{\circ}$. This result is consistent with both the planet's orbit being aligned or highly misaligned with the spin-axis of its host star. Our measurement independently agrees with, but is far less precise than observations carried out on other instruments around the same time that measure a low obliquity orbit for the planet. AU\,Mic is the youngest exoplanetary system for which the projected spin-orbit angle has been measured, making it a key data point in the study of the formation and migration of exoplanets -- particularly given that the system is also host to a bright debris disk.
\end{abstract}


\section{Introduction} \label{sec:intro}

Prior to the discovery of the first planets orbiting other stars, our only laboratory for the study of planet formation was the Solar System. Whilst our planetary system holds a great wealth of information on the way in which planetary systems form and evolve \citep[as described in detail in the recent review by][]{SSRev}, it represents just one possible outcome of that process. For that reason, the discovery of the first exoplanets \citep[e.g.][]{GammaCeph,LathamsWorld,PSRblah,51Peg} led to a revolution in our understanding of planet formation -- giving our first insight into the true diversity of outcomes for the planet formation process.

One of the most startling discoveries of the exoplanet era has been that of the so-called `hot Jupiters' -- giant planets moving on orbits that almost skim the surface of their host stars \citep[e.g.][]{HJ1,HJ2,HJ3}. A number of mechanisms have been proposed to explain the migration of the hot Jupiters - ranging from planet-planet scattering \citep[e.g.][]{Scatter2,Scatter1} to interaction with the material within the protoplanetary disk \citep[e.g.][]{Disk1,Disk2,Disk3,Disk4} -- and even orbital excitation by a distant companion of their host star, followed by a process of tidal circularisation, locking the planet's orbit in at the distance of its periastron passage \citep[e.g.][]{Kozai1,Kozai2,Kozai3,Kozai4}. For a review of the different processes that may play a role in the formation of hot Jupiters, we direct the interested reader to \citet{DawJJ}.

Distinguishing between these planet migration mechanisms is currently a leading goal of exoplanetary science. Each migration mechanism would result in a dramatically different planetary system. A variety of observational methods can come together to identify which have been active in a given system hosting a short-period planet. Of particular interest here are observations of the `Rossiter-McLaughlin effect' \citep{RM1,RM2}, which allows the sky-projected inclination (obliquity) of the planet's orbit with respect to the plane of its host star's equator to be accurately determined. Such observations have revealed a significant population of strongly misaligned exoplanets \citep[e.g.][]{MA1,MA2,MA3}, including planets moving on retrograde orbits \citep[e.g.][]{retro1,retro2}. 

In this context, it is particularly interesting to study planets that have only recently formed, or are in the process of formation and migration. The transiting planet orbiting AU~Microscopii, AU~Mic b, is a particularly interesting target in this regard. AU~Mic is a young \citep[23~$\pm$~3~Myr,][]{2014MNRAS.445.2169M}, nearby \citep[$d = 9.725~\pm~0.005~$pc, ][]{2018yCat.1345....0G} M-type star \citep[M1~V,][]{1989ApJS...71..245K}, surrounded by a substantial, spatially resolved debris disk \citep[e.g. ][]{2004Sci...305.1442L,2004Sci...303.1990K,2013ApJ...762L..21M}. It is known to host at least one planet - AU Mic b, which transits its host every 8.46 days \citep{plavchan20}. Additional planets in the system are suspected from multi-wavelength radial velocity measurements \citep{plavchan20} and tentative evidence of disk sub-structure at millimetre wavelengths \citep{2019ApJ...875...87D}.

The disk around AU\,Mic has been imaged at a wide range of wavelengths, revealing its orientation and extent \citep[e.g.,][]{2004Sci...303.1990K,2013ApJ...762L..21M,2015ApJ...811..100M,2017MNRAS.470.3606H}. Overall, the disk architecture is a single broad and co-planar belt oriented edge-on, extending out to 210 au in scattered light \citep{2004Sci...303.1990K}, with the dust-producing planetesimal belt located around 40 au from the star \citep{2013ApJ...762L..21M}. Collisonal modelling of the disk suggested that the disk should be dynamically cold \citep{2015A&A...581A..97S}; recent high angular resolution ALMA observations of the disk have resolved its vertical extent, finding it to be vertically thin and unstirred, ruling out the influence of planets more massive than a few times the Earth on its dynamics. The disk appears to be well-aligned with AU\,Mic's stellar equator \citep{2014MNRAS.438L..31G}, therefore planetary companions might also be co-aligned. Under the assumption that the disk is aligned with the equatorial plane of AU~Mic, measurement of AU\,Mic b's obliquity offers a fascinating insight into the formation and evolution of a new hot Neptune system. 

In that light, we present herein the results of Rossiter-McLaughlin observations of three spectroscopic transits of AU\,Mic\,b, observed by the \textsc{Minerva}-Australis array \citep{addison2019}.  In Section~\ref{sec:observations}, we describe the radial velocity observations, Section~\ref{sec:RM_analysis} presents the Rossiter-McLaughlin analysis and results, and we give our conclusions in Section~\ref{sec:Discussion}. 

This work is complemented by three additional studies \citep{AUM1,AUM2,AUM3}, each of which investigated the transits of AU~Mic b that occurred in early 2020. Those papers were submitted in parallel to this work, and represent a suite of new observations that describe the first ever studies of the orbital alignment of such a young and newly formed exoplanet.

\section{Observations and Data Reduction}\label{sec:observations}
We carried out the spectroscopic observations of three AU\,Mic\,b transits using the {\textsc{Minerva}}-Australis facility \citep{2018arXiv180609282W,addison2019,2020arXiv200107345A}. {\sc {\textsc{Minerva}}}-Australis consists of an array of four independently operated 0.7\,m CDK700 telescopes situated at the Mount Kent Observatory in Queensland, Australia \citep{addison2019}. Each telescope simultaneously feeds stellar light via fiber optic cables to a single KiwiSpec R4-100 high-resolution ($R=80,000$) spectrograph \citep{2012SPIE.8446E..88B} with wavelength coverage from 480 to 620\,nm.

We observed two partial transits on 31 May 2019 and 17 June 2019 and one full transit on 18 September 2019. For the 31 May 2019 transit observation, we started observing AU\,Mic at 13:19\,UT just prior to mid transit at an airmass of 2.25 using telescopes 3 and 5 (corresponding to fiber numbers 3 and 5, respectively) in the {\textsc{Minerva}}-Australis array. Exposure times for these observations were set to 1200\,s with a duty cycle of 1250\,s (including all overheads), providing a signal-to-noise ratio between $\sim17$ and $\sim42$ per resolution element at $\sim550$\,nm for each of the fibers. These observations continued until 17:50\,UT, providing six in-transit and eight out-of-transit radial velocities.

For the 17 June 2019 transit observation, observing started at 12:14\,UT during transit egress and at an airmass of $\sim2$ that decreased throughout the night, and lasted for 4.25\,hr until 18:34\,UT. Exposure times were set to 900\,s for this observation (total cadence of 950s) using telescopes 1, 3, and 4 (fibers 4, 3, and 6, respectively) in the {\textsc{Minerva}}-Australis array, yielding a signal-to-noise ratio between $\sim13$ and $\sim24$ per resolution element in each of the three fibers. A total of six in-transit radial velocities were obtained for this transit observation.

On 18 September 2019, we carried out a full transit observation of AU\,Mic starting at 11:33\,UT ($\sim1$\,hr before transit ingress) and continued until 16:04\,UT ($\sim15$\,m after transit egress). The airmass ranged from $\sim1$ to $\sim2.9$ throughout the observations. We set the exposures to 900\,s, yielding a signal-to-noise ratio between $\sim13$ and $\sim24$ per resolution element in each of the fibers and 11 in-transit and six out-of-transit radial velocities using telescopes 1, 3, and 4. One observation taken during transit egress (BJD 2458745.147465278) was affected by low signal-to-noise likely as a result of poor guiding in all three telescopes and has been excluded from the analysis.

Radial velocities for the observations are derived for each telescope by using the least-squares technique of \citet{anglada2012}, where the template being matched is the mean out-of-transit spectrum of each telescope. Spectrograph drifts are corrected for using simultaneous Thorium-Argon (ThAr) arc lamp observations. The radial velocities from each telescope are given in Table~\ref{table:unbinned} for the transits observed on 31 May, 17 June, and 18 September, respectively, and labeled by their fiber number. For the Rossiter-McLaughlin analysis, we binned together the radial velocities taken at the same time with each individual telescope as one radial velocity. By binning the data across {\textsc{Minerva}}-Australis telescopes, we avoid needing to account for the systematics common to all the {\textsc{Minerva}}-Australis radial velocities. We achieve a median internal precision of $16$\,\mos\, with the binned radial velocities and they are given in Table~\ref{table:binned}. We also measured the projected stellar rotational velocity, $v\sin i$, of AU\,Mic by fitting a rotationally broadened Gaussian \citep{2005oasp.book.....G} to a least-squares deconvolution profile \citep{1997MNRAS.291....1D} obtained from the sum of all the spectral orders from the combined highest S/N {\sc {\textsc{Minerva}}}-Australis spectra of the star. The resulting $v\sin i$ is $12.2\pm0.7$\,\kms.

\startlongtable
\begin{deluxetable}{cccc}
\tabletypesize{\scriptsize}
\tablecaption{{\textsc{Minerva}}-Australis Radial Velocities for AU\,Mic for the three Transit Observations\label{table:unbinned}}
\tablehead{
\colhead{Time} & \colhead{Velocity} & \colhead{Uncertainty} & \colhead{Fiber}\\
\colhead{[BJD]} & \colhead{[\mos]} & \colhead{[\mos]} & \colhead{}}
\startdata
  \multicolumn4c{31 May 2019 transit observations} \\
  2458635.055428  &   -4787  &   22   & 3  \\
  2458635.055428  &   -4882  &   17   & 5  \\
  2458635.069884  &   -4849  &   26   & 3  \\
  2458635.069884  &   -4698  &   20   & 5  \\
  2458635.084352  &   -4939  &   26   & 3  \\
  2458635.084352  &   -4880  &   22   & 5  \\
  2458635.098808  &   -4864  &   26   & 3  \\
  2458635.098808  &   -4758  &   24   & 5  \\
  2458635.113275  &   -4921  &   26   & 3  \\
  2458635.127743  &   -4838  &   26   & 3  \\
         :        &     :    &   :    & : \vspace{4pt} \\
  \hline
  \multicolumn4c{17 June 2019 transit observations} \\
  2458652.009942  &   -4660  &   26   & 3  \\
  2458652.009942  &   -4641  &   26   & 4  \\
  2458652.009942  &   -4556  &   24   & 6  \\
  2458652.020926  &   -4592  &   26   & 3  \\
  2458652.020926  &   -4653  &   26   & 4  \\
  2458652.031921  &   -4593  &   26   & 3  \\
  2458652.031921  &   -4586  &   26   & 4  \\
  2458652.031921  &   -4633  &   25   & 6  \\
  2458652.042905  &   -4580  &   24   & 3  \\
  2458652.042905  &   -4530  &   26   & 4  \\
         :        &     :    &   :    & : \vspace{4pt} \\
  \hline
  \multicolumn4c{18 Sept. 2019 transit observations} \\
  2458744.981713  &   -5208  &   31   & 6  \\
  2458744.981713  &   -5237  &   24   & 4  \\
  2458744.981713  &   -5368  &   30   & 3  \\
  2458744.993333  &   -5193  &   28   & 4  \\
  2458744.993333  &   -5287  &   28   & 3  \\
  2458744.993333  &   -5312  &   38   & 6  \\
  2458745.002963  &   -5250  &   25   & 4  \\
  2458745.002963  &   -5253  &   25   & 3  \\
  2458745.002963  &   -5255  &   32   & 6  \\
  2458745.015000  &   -5201  &   38   & 6  \\
         :        &     :    &   :    & : \vspace{4pt} \\
\enddata
\tablecomments{Table~\ref{table:unbinned} is published in its entirety in the machine-readable format. A portion is shown here for guidance regarding its form and content.}
\end{deluxetable}

\startlongtable
\begin{deluxetable}{ccc}
\tablecaption{Binned {\textsc{Minerva}}-Australis Radial Velocities for the Three Transit Observations\label{table:binned}}
\tablewidth{0pt}
\tablehead{
\colhead{Time} & \colhead{Radial velocity} & \colhead{Uncertainty} \\
\colhead{[BJD]} & \colhead{[\mos]} & \colhead{[\mos]}}
\startdata
  \multicolumn3c{31 May 2019 transit observations} \\
  2458635.055428  &   -4845.0  &   14.0  \\
  2458635.069884  &   -4754.0  &   16.0  \\
  2458635.084352  &   -4905.0  &   17.0  \\
  2458635.098808  &   -4806.0  &   18.0  \\
  2458635.113275  &   -4921.0  &   26.0  \\
  2458635.127743  &   -4838.0  &   26.0  \\
  2458635.142199  &   -4852.0  &   25.0  \\
  2458635.156667  &   -4835.0  &   24.0  \\
  2458635.171123  &   -4848.0  &   25.0  \\
  2458635.185590  &   -4852.0  &   25.0  \\
         :        &      :     &    :    \vspace{4pt} \\
           \hline
  \multicolumn3c{17 June 2019 transit observations} \\
  2458652.009942  &   -4615.0  &   15.0  \\
  2458652.020926  &   -4622.0  &   19.0  \\
  2458652.031921  &   -4605.0  &   15.0  \\
  2458652.042905  &   -4532.0  &   15.0  \\
  2458652.053900  &   -4561.0  &   15.0  \\
  2458652.064884  &   -4528.0  &   15.0  \\
  2458652.075880  &   -4569.0  &   15.0  \\
  2458652.086863  &   -4526.0  &   15.0  \\
  2458652.097859  &   -4541.0  &   15.0  \\
  2458652.108843  &   -4514.0  &   14.0  \\
         :        &      :     &    :    \vspace{4pt} \\
           \hline
  \multicolumn3c{18 Sept. 2019 transit observations} \\
  2458744.981713  &   -5266.0  &   16.0  \\
  2458744.993333  &   -5255.0  &   18.0  \\
  2458745.002963  &   -5253.0  &   16.0  \\
  2458745.015000  &   -5251.0  &   18.0  \\
  2458745.025995  &   -5246.0  &   21.0  \\
  2458745.038738  &   -5260.0  &   22.0  \\
  2458745.049850  &   -5265.0  &   18.0  \\
  2458745.060833  &   -5284.0  &   17.0  \\
  2458745.070613  &   -5273.0  &   17.0  \\
  2458745.084120  &   -5280.0  &   18.0  \\
         :        &      :     &    :    \vspace{4pt} \\
\enddata
\tablecomments{Table~\ref{table:binned} is published in its entirety in the machine-readable format. A portion is shown here for guidance regarding its form and content.}
\end{deluxetable}

\section{Rossiter--McLaughlin Analysis}\label{sec:RM_analysis}
We determined the sky-projected spin-orbit angle ($\lambda$) for AU\,Mic\,b from spectroscopic observations of the Rossiter-McLaughlin effect using a custom \texttt{Python} script that incorporates the \citet{2011ApJ...742...69H} Rossiter-McLaughlin model and the \texttt{batman} photometric transit model \citep{2015PASP..127.1161K}. For this analysis, we performed the fit on the three Rossiter-McLaughlin transit observations simultaneously using radial velocities binned by telescope. To sample the posterior distributions, we used the \texttt{emcee} Markov Chain Monte Carlo (MCMC) package \citep{2013PASP..125..306F}.

AU\,Mic is an extremely young star that displays significant chromospheric activity \citep[i.e., spots, plages, and flares. See e.g.][]{IbanezBustos2019,MacGregor2020} causing elevated levels of stellar signal in the radial velocity data. The {\textsc{Minerva}}-Australis observations of the three transits show strong evidence for stellar activity in the form of positive and negative slopes in the radial velocities. Given that the rotation period of AU\,Mic is 4.8\,days, considerably longer than the $\sim4$\,hr transit duration and the length of each transit observation, the changes in the spectrum due to photospheric features are expected to be relatively smooth and stable. The potential exceptions are flare events or large star spots on the visible surface during a transit, both of which can alter the observed Rossiter-McLaughlin effect signal (star spots discussed in \ref{sec:spots}). Therefore, the stellar activity signal should be mostly accounted for and removed from the radial velocity data using linear trends or second order polynomials. To account for the radial velocity activity (as well as the planetary) signal, we have trialed in our model a hybrid linear slope and a second order hybrid polynomial. At each step in the MCMC, a least squares minimization determines the linear (or polynomial) parameters to set the baseline of each of the three transit observations, following the hybrid polynomial procedure implemented in \texttt{Allesfitter} \citep{allesfitter-code,allesfitter-paper}. The hybrid polynomial model represents a more conservative approach to accounting for the effects of stellar activity compared to the hybrid linear model and as such, we adopt the results from that model as our preferred solution.


Table~\ref{table:results} lists the priors, the $1\sigma$ uncertainties, and the prior type of each parameter used in the fitting of the radial velocities acquired during the three transit events. The results of the MCMC analysis and the solutions for $\lambda$ and $v\sin i_{\star}$ for both the hybrid linear slope and polynomial activity models are also given in Table~\ref{table:results}.

For the Rossiter-McLaughlin analysis, we imposed Gaussian priors on the model parameters from the reported values in \citet{plavchan20} on the planet-to-star radius ratio ($R_{P}/R_{\star}$), mid-transit epoch ($T_{0}$), orbital period ($P$), inclination angle ($I$), semi-major axis to star radius ratio ($a/R_{\star}$), and an inflated 5\,$\sigma$ (weak prior) on $v\sin i_{\star}$ of $12.2\pm3.5$\kms, as measured from the {\textsc{Minerva}}-Australis spectra. Uniform priors are used on the quadratic limb-darkening coefficients ($q_{1}$ and $q_{2}$) with boundaries between 0 and 1 and starting values of 0.47 and 0.40 based on interpolated values from look-up tables in \citet{2011A&A...529A..75C} using the Johnson V-band and stellar parameters close to those for AU Mic. We have included independent radial velocity jitter terms for each transit observation ($jit_1$, $jit_2$, and $jit_3$ for the 31 May 2019, 17 June 2019, and 18 September 2019 observations, respectively) using uniform priors bounded between 0.1 and 6.9\,\mos\, in natural log space. A uniform prior is also used for $\lambda$ which is bounded between $-180$ and $+180$\,degrees. We fixed the orbital eccentricity ($e$) to 0, the adopted solution in \citet{plavchan20}, and the radial velocity semi-amplitude ($K$) to 0, since both the hybrid linear and polynomial activity models will account for the stellar activity and the small planetary signal.

\begin{figure*}
\gridline{\fig{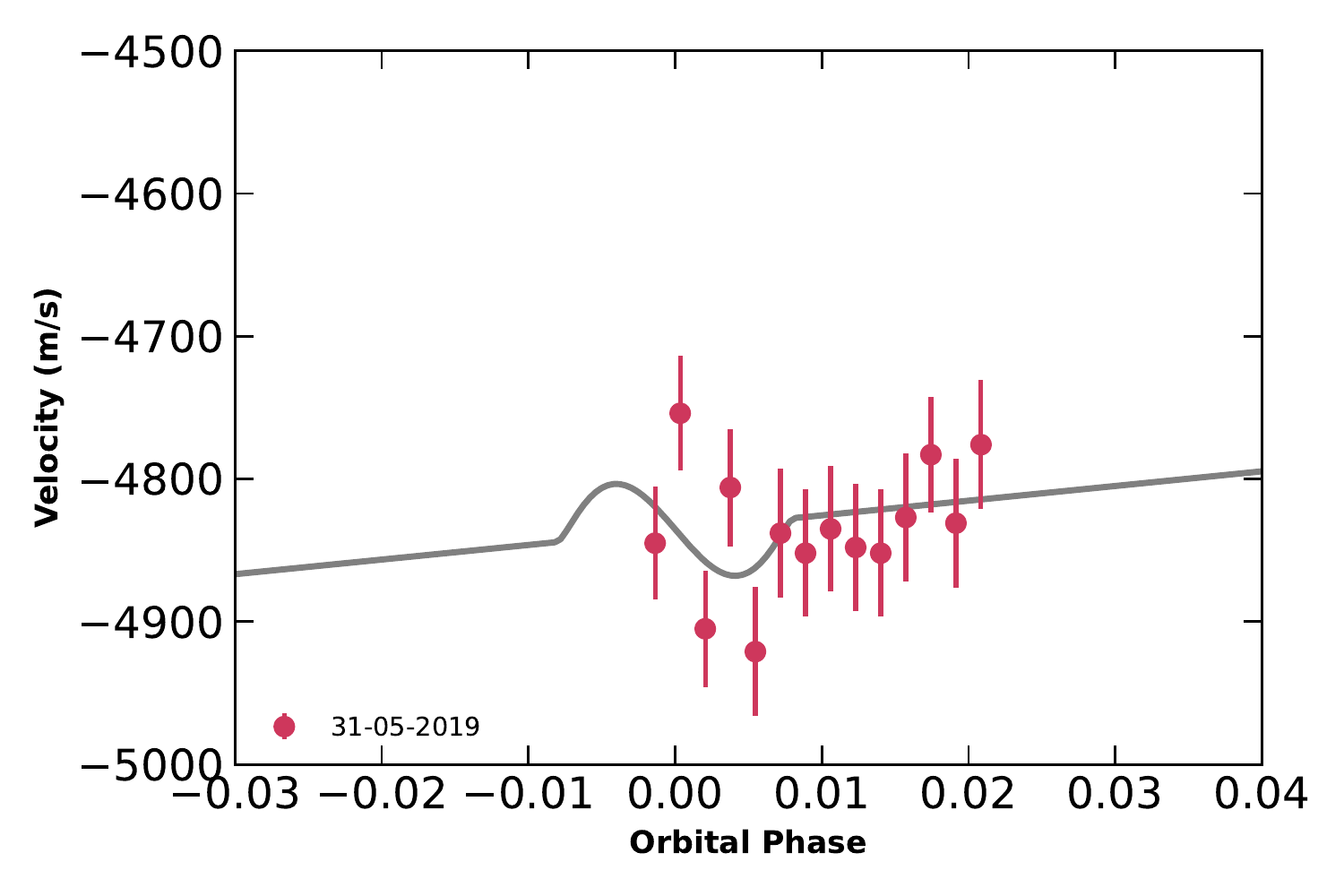}{0.48\textwidth}{(a)}
          \fig{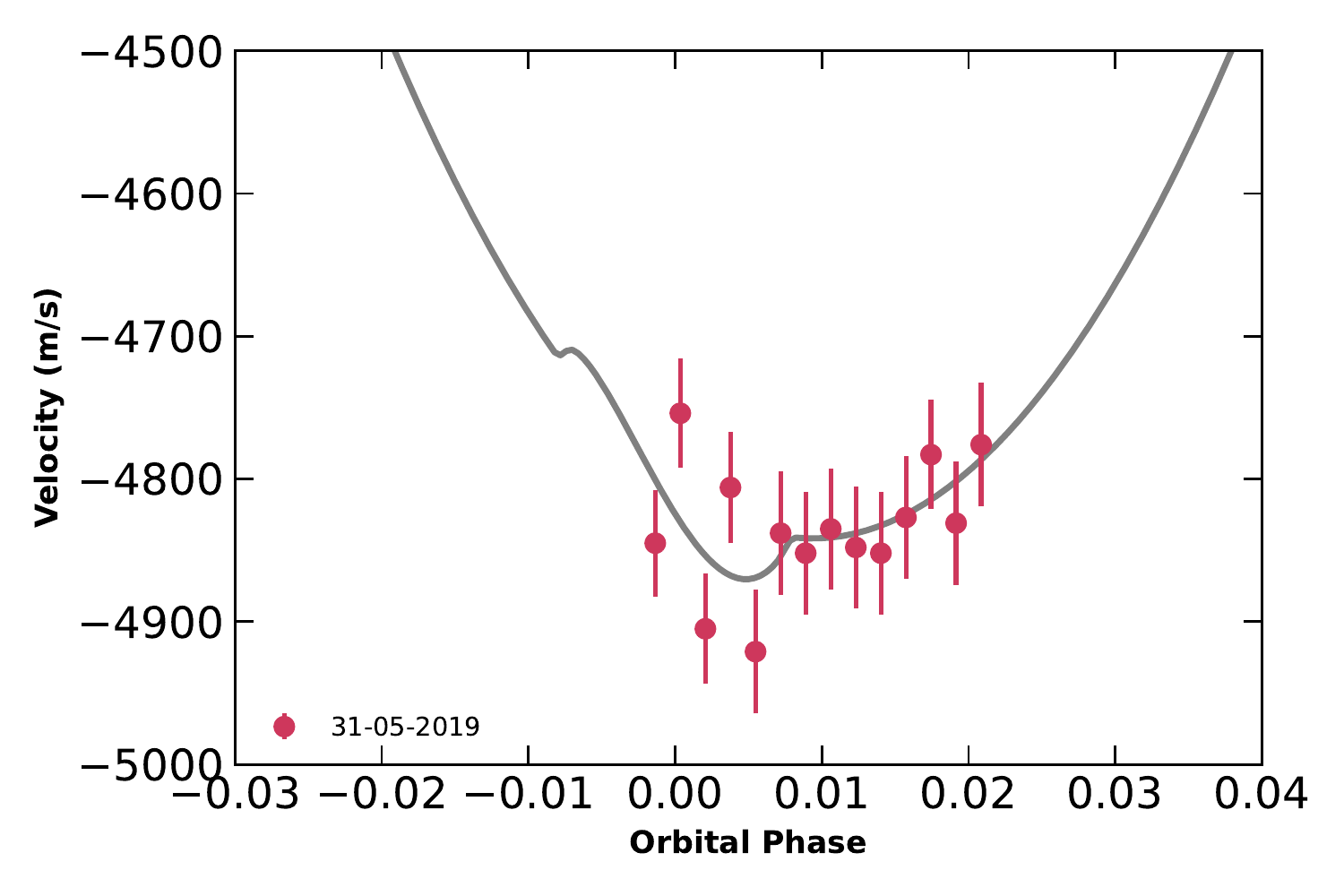}{0.48\textwidth}{(b)}
          }
\gridline{\fig{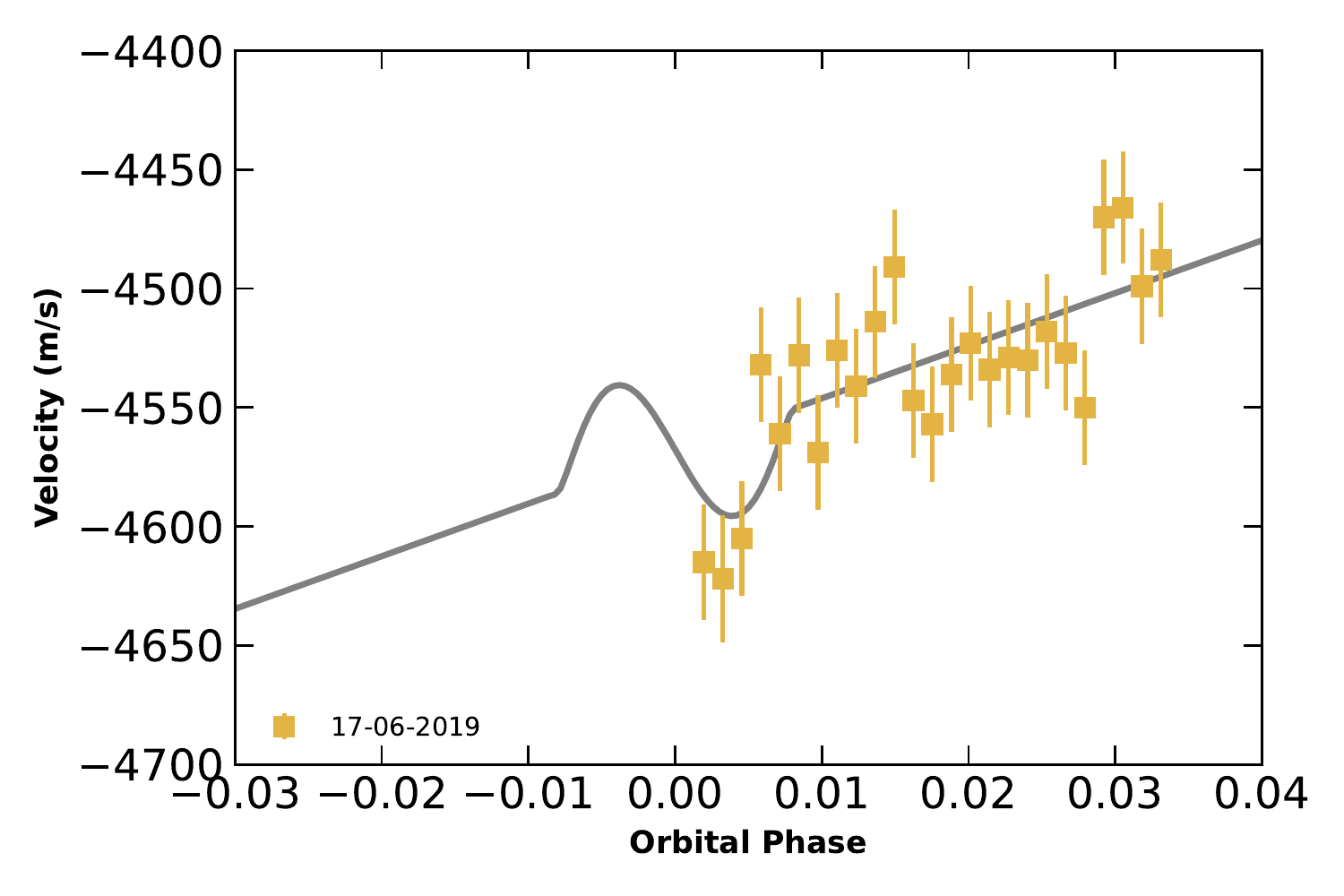}{0.48\textwidth}{(c)}
          \fig{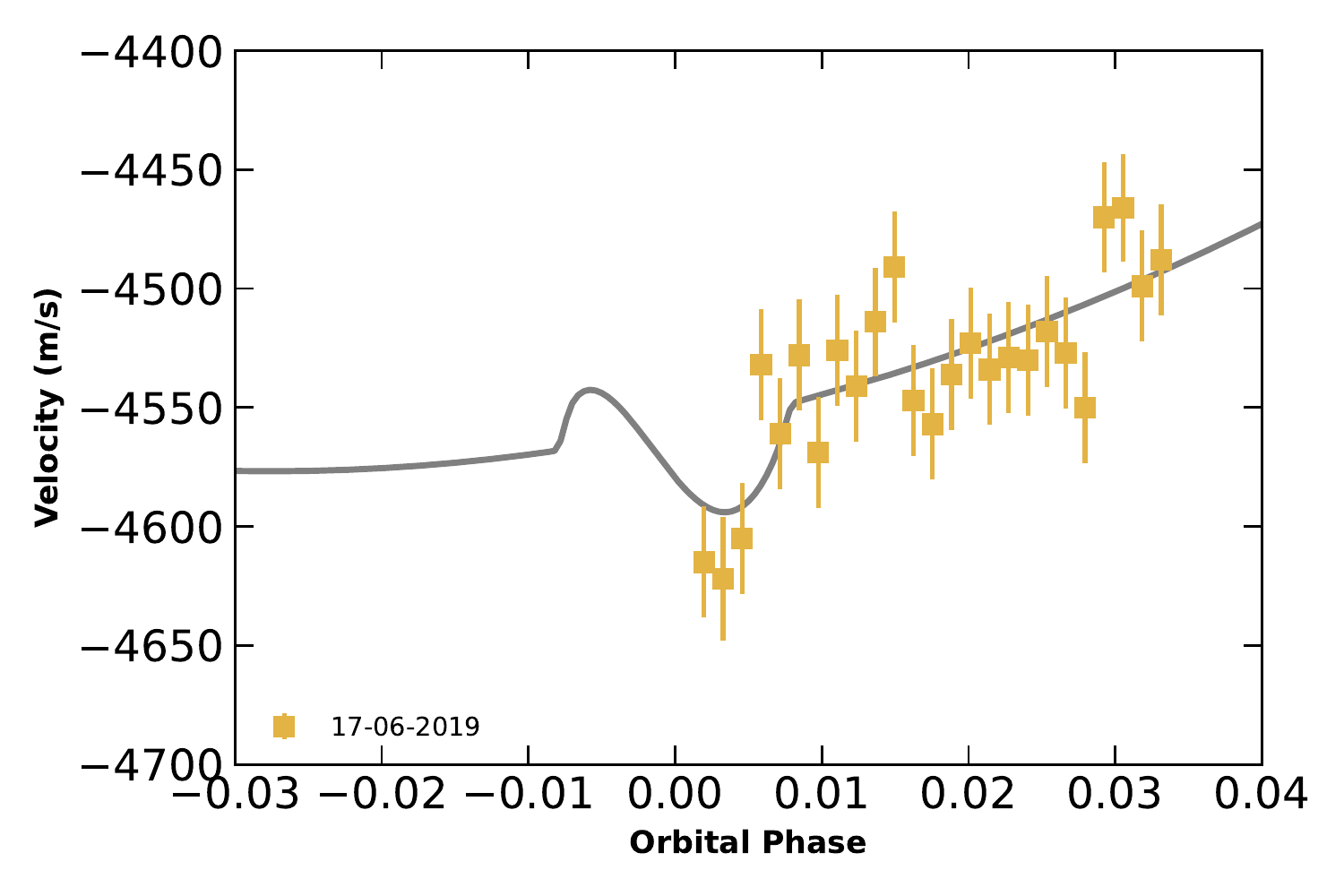}{0.48\textwidth}{(d)}
          }
\gridline{\fig{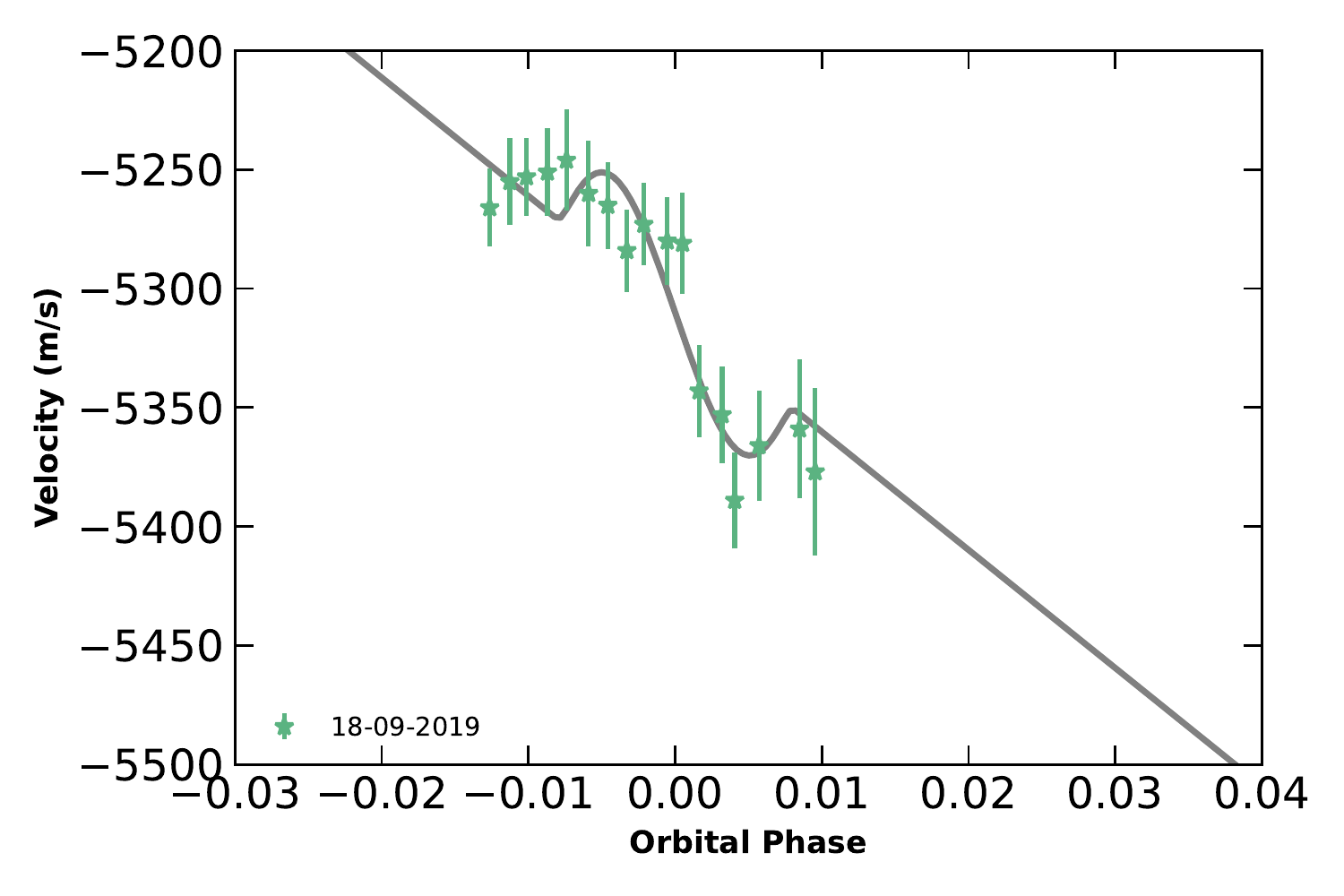}{0.48\textwidth}{(e)}
          \fig{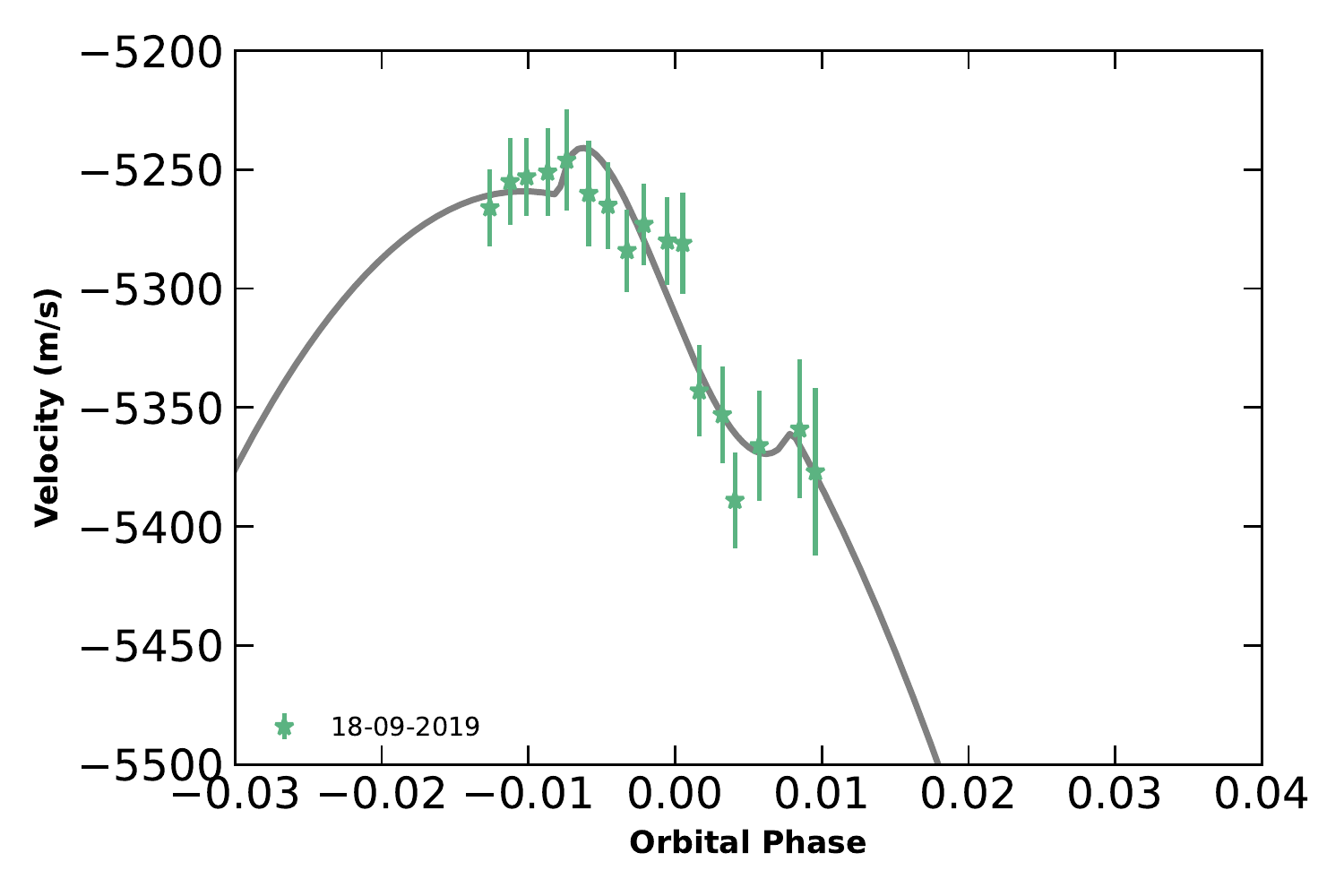}{0.48\textwidth}{(f)}
          }
\caption{Telescope binned spectroscopic radial velocities of three AU\,Mic\,b transits, plotted as a function of orbital phase with the best fitting Rossiter-McLaughlin model + stellar activity model (hybrid linear plotted on the left panels and second order hybrid polynomial plotted on the right panels as the gray line). The transit observation on 31 May 2019 is shown in (a) and (b), 17 June 2019 is shown in (c) and (d), and 18 September 2019 is shown in (e) and (f). The filled red circles, orange squares, and teal stars with error bars are the binned radial velocities obtained in this work on the 31 May 2019, 17 June 2019, and 18 September 2019, respectively.
\label{fig:individual_transits}}
\end{figure*}

\begin{figure*}
\plotone{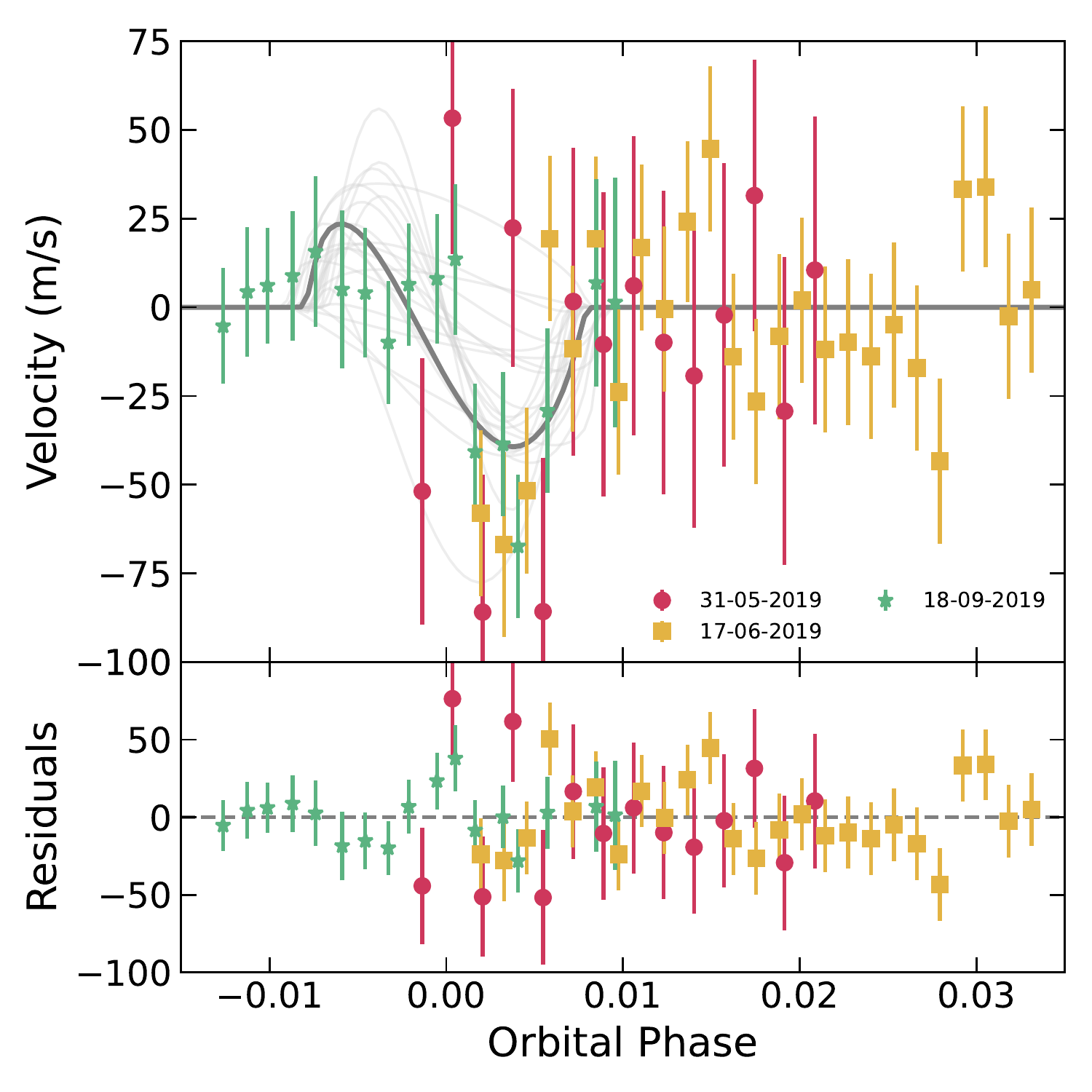}
\caption{Telescope binned spectroscopic radial velocities of AU\,Mic, phased to a single transit, plotted as a function of phase with the best-fit median Rossiter-McLaughlin model (solid opaque gray line), 20 Rossiter-McLaughlin models drawn from the posterior (translucent gray lines), and corresponding residuals (from the best-fit model). The second order hybrid polynomial (preferred solution) for each transit observation has been removed from the radial velocities. The filled red circles, orange squares, and teal stars with error bars are the binned radial velocities from the 31 May 2019, 17 June 2019, and 18 September 2019 transit observations, respectively.
\label{fig:combined_transits}}
\end{figure*}

\begin{figure*}
\plotone{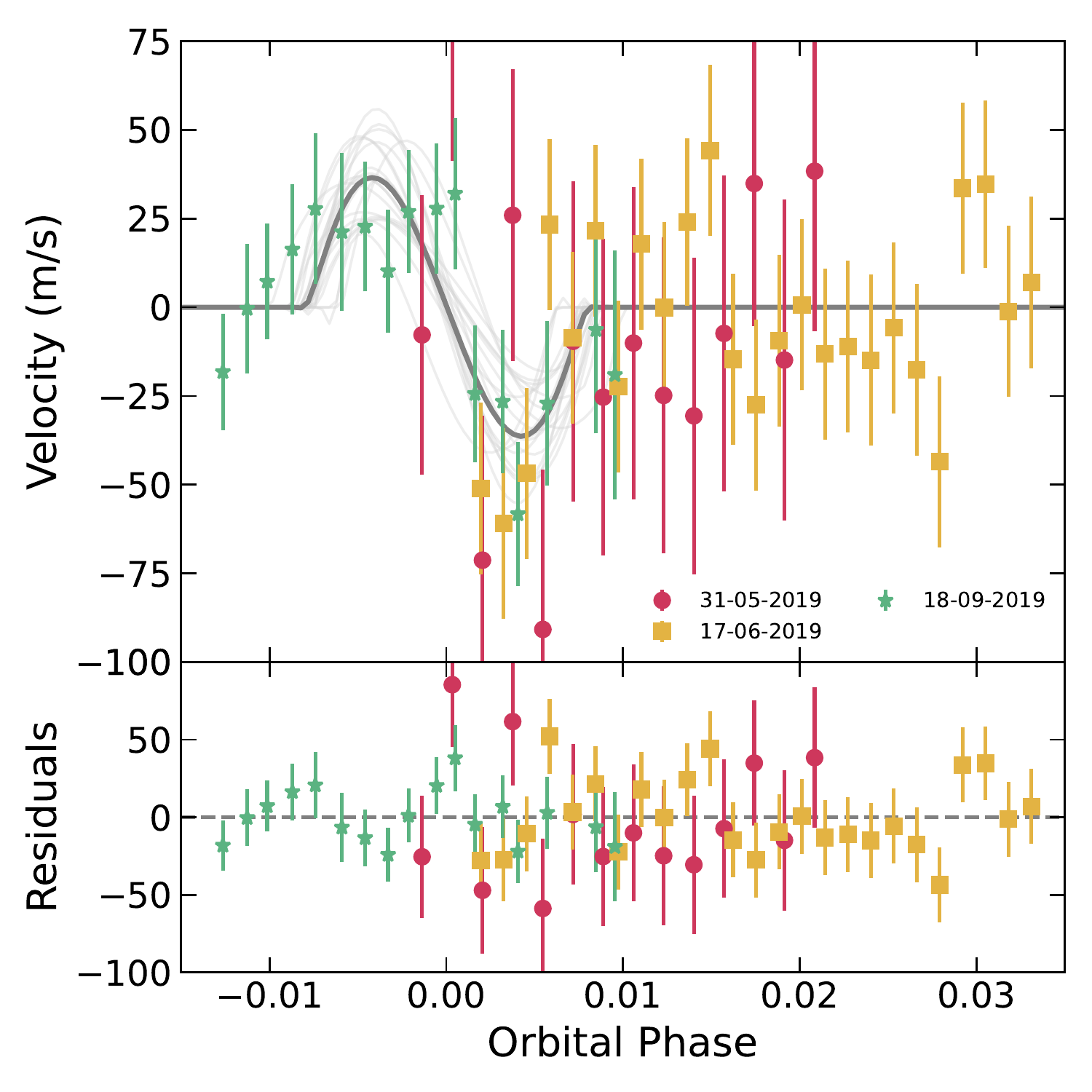}
\caption{Same as Figure~\ref{fig:combined_transits}, but the stellar activity has been removed using a hybrid linear model.
\label{fig:combined_transits_linear}}
\end{figure*}

The MCMC was run with 100 walkers, 20,000 total steps for each walker (of which the first 500 were discarded as burn-in), and the chains were thinned by a factor of 10, resulting in a total of 195,000 samples. All chains were greater than $30\times$ their auto-correlation lengths, indicating the MCMC had reached convergence. The observations and the resulting best-fit models (Rossiter-McLaughlin + hybrid linear or polynomial) for each individual transit are shown in Figure~\ref{fig:individual_transits}. Figures~\ref{fig:combined_transits} and \ref{fig:combined_transits_linear} show the radial velocities phased to a single transit with either the hybrid polynomial or linear trend, respectively, removed from the data, over plotted with the best-fit Rossiter-McLaughlin model, and 20 model samples randomly drawn from the posterior. Figures~\ref{fig:posterior} and \ref{fig:posterior_linear} in the Appendix are the resulting posterior distribution corner plots for the hybrid polynomial and linear models, respectively. We find the best-fit projected spin-orbit angle is $\lambda = -2{^{+27}_{-26}}^{\circ}$ with the hybrid linear activity model and $\lambda = 47{^{+26}_{-54}}^{\circ}$ using the preferred second order hybrid polynomial model. The linear activity model provides a more precise measurement of $\lambda$ compared with the more conservative second order polynomial model, which is likely the result of the polynomial removing some of the Rossiter-McLaughlin signal from the radial velocities.

The lack of full transit coverage for the radial velocities obtained on the 31 May 2019 and 17 June 2019 as well as the lack of sufficient out-of-transit baseline for the full transit observation obtained on the 18 September 2019 renders the complete removal of the stellar activity signal in the radial velocities challenging and a precise measurement of the system`s sky-projected obliquity impossible. The low-precision obliquity measurement reported here is consistent with the orbit of AU\,Mic\,b being well-aligned or highly misaligned, but seems to suggest that a retrograde orbit ($|\lambda|\geq135^{\circ}$) is unlikely at $>3\sigma$.

Our result is consistent with, but less precise than, the low-obliquity measured for AU Mic b by other studies submitted in parallel with this paper. These include \citet{AUM1} result of $\lambda = -2.96{^{+10.44}_{-10.30}}^{\circ}$ from Rossiter-McLaughlin observations taken with the ESPRESSO spectrograph on the Very Large Telescope array, \citet{AUM2} result of $\lambda = -4.7{^{+6.8}_{-6.4}}^{\circ}$ from Doppler tomography observations using the IRD spectrograph on the Subaru telescope, and the \citet{AUM3} result of $\lambda = 0{^{+18}_{-15}}^{\circ}$ from Rossiter-McLaughlin observations using the SPIRou spectrograph on the Canada-France-Hawaii Telescope.

\subsection{Star Spots}\label{sec:spots}
A study by \citet{2018A&A...619A.150O} has shown that star spots can lead to variations in the shape of the Rossiter-McLaughlin signal observed in the radial velocity data, even when no spot crossing events occur during a transit. These variations in the observed signal can lead to biased results for the measured spin-orbit angle of up to $\sim42^{\circ}$. With this in mind and given AU Mic`s extreme youth ($\sim23$\,Myr) and chromospheric activity (as observed by \textit{TESS}), we have investigated the potential effects of star spots on the Rossiter-McLaughlin radial velocity signal to determine the impact that they may have on the results reported in this work. However, due to the lack of simultaneous photometry obtained during the {\textsc{Minerva}}-Australis Rossiter-McLaughlin observations as well as the time gap between those observations and when \textit{TESS} observed AU Mic between 2018 July 25--August 22 ($\sim280$\,days), we have decided not to include a star spot model into the Rossiter-McLaughlin analysis given the high likelihood of the star spots significantly evolving during the time gap.

To highlight the influence of stellar intrinsic variability on the Rossiter-McLaughlin signature, we modeled several different single spot configurations on the stellar surface during a transit of AU Mic b using the spot model of Heitzmann et al., in prep. The star was modeled as a disk of uniform brightness on a pixel grid to which the following elements were added: (i) a quadratic limb darkening law; (ii) a single spot located at a different co-latitude $\theta$ and longitude $\phi$ on the stellar surface for each of the model configurations and with a radius yielding a $\sim5$\,\% variation in flux over a stellar rotation (as observed by \textit{TESS}, see \citealt{plavchan20}); and (iii) a transiting planet on an aligned orbit ($\lambda = 0^{\circ}$). The values applied to the stellar and planetary parameters in this model are fixed to the prior values given in Table~\ref{table:results} that were used for the Rossiter-McLaughlin analysis carried out in this work.

\begin{figure*}
\gridline{\fig{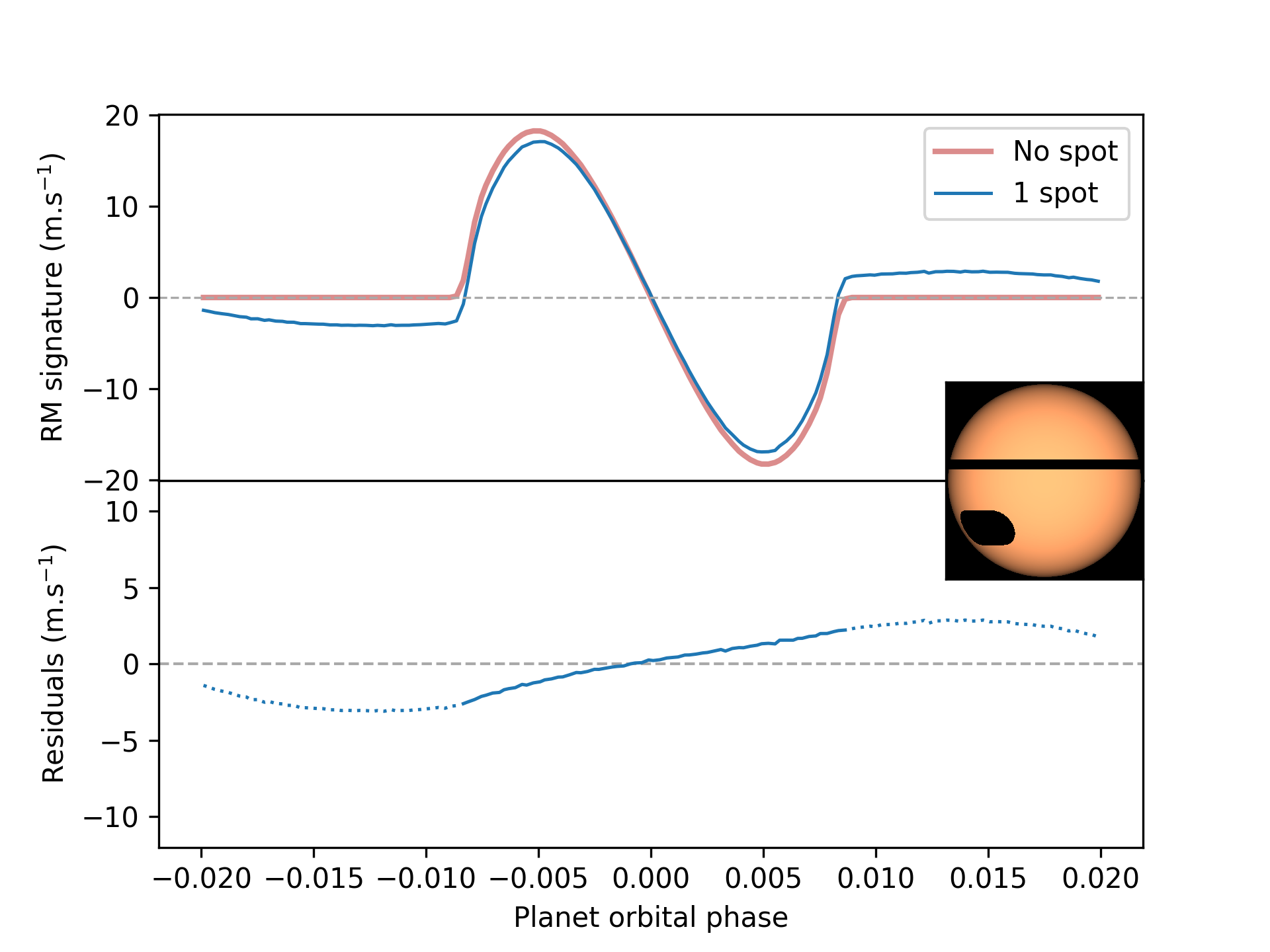}{0.48\textwidth}{(a)}
          \fig{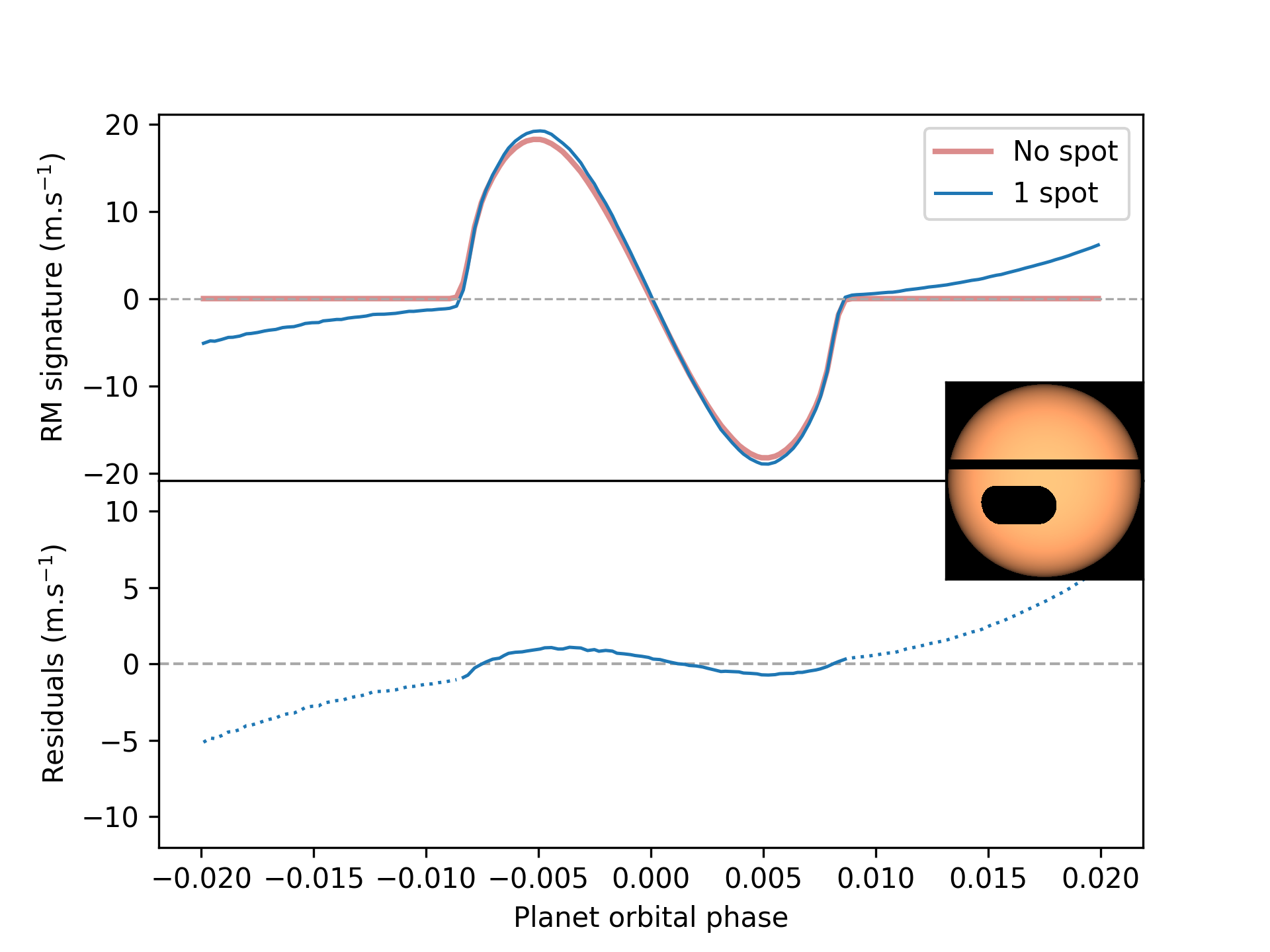}{0.48\textwidth}{(b)}
          }
\gridline{\fig{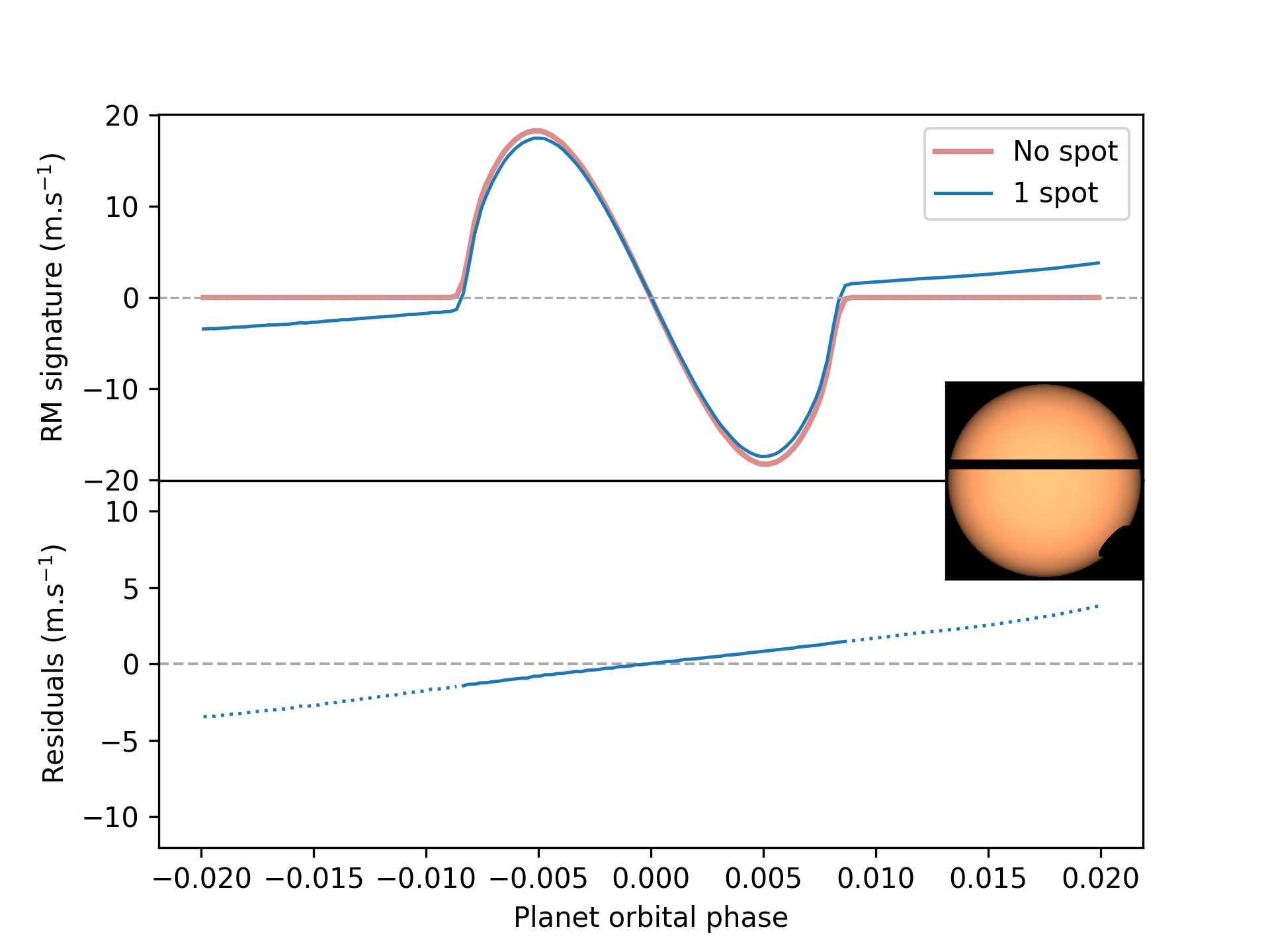}{0.48\textwidth}{(c)}
          \fig{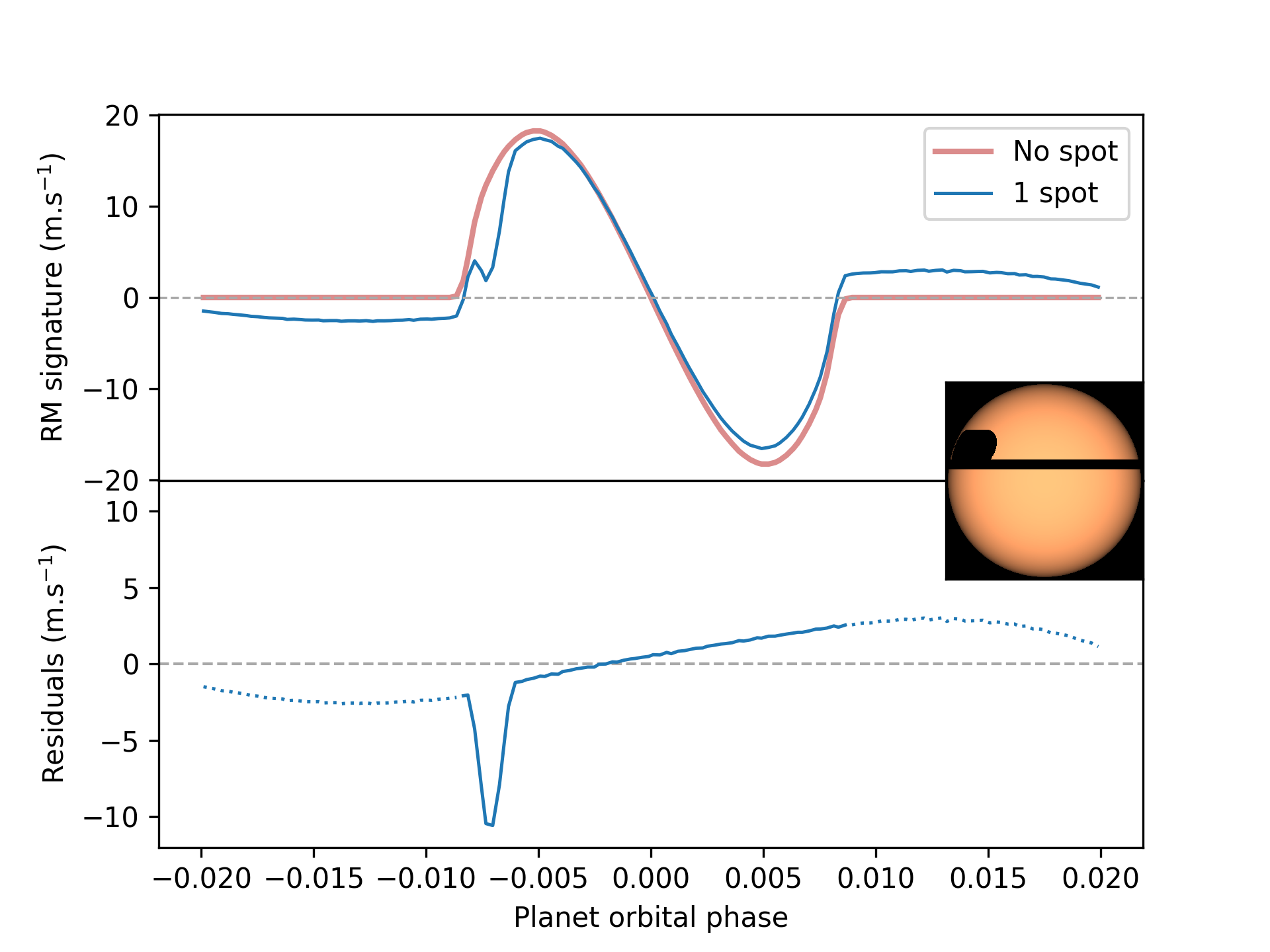}{0.48\textwidth}{(d)}
          }
\caption{Simulations of AU Mic b's transit with four different single star spot configurations. In each figure, the top panel shows the Rossiter-McLaughlin signature produced for the planet assumed to be on an aligned orbit ($\lambda = 0^{\circ}$) for an un-spotted stellar surface (red line) and with a single star spot, detrended with a second order polynomial (blue line). The bottom panel shows the residuals between the Rossiter-McLaughlin only model and Rossiter-McLaughlin + star spot model (red curve - blue curve). Shown on the right of each plot are stacked images of the planet (thin horizontal dark streak) and the spot travelling across the surface (thicker dark patch) during the transit event. (a) starting spot position: $\theta_{\mathrm{spot}}$ = $120^{\circ}$ and $\phi_{\mathrm{spot}}$ = $300^{\circ}$, (b) starting spot position: $\theta_{\mathrm{spot}}$ = $105^{\circ}$ and $\phi_{\mathrm{spot}}$ = $330^{\circ}$, (c) starting spot position: $\theta_{\mathrm{spot}}$ = $130^{\circ}$ and $\phi_{\mathrm{spot}}$ = $70^{\circ}$ and (d) starting spot position: $\theta_{\mathrm{spot}}$ = $70^{\circ}$ and $\phi_{\mathrm{spot}}$ = $290^{\circ}$. A spot crossing event occurs, resulting in a temporary $\sim-10$\,\mos\, anomaly in the Rossiter-McLaughlin signature.}
\label{fig:simulated_R-M}
\end{figure*}

Four of the simulations are shown in Figure~\ref{fig:simulated_R-M} to highlight the impact of a star spot on the Rossiter-McLaughlin signature. From the trialed star spot configurations, we find that unless the planet directly crosses over a spot, the effect on the resulting radial velocity Rossiter-Mclaughlin signal (blue line on Figure~\ref{fig:simulated_R-M}) is small ($< 5$\,\mos\, for all spot configurations except for spot crossing events as shown in Figure~\ref{fig:simulated_R-M}\,(d)). We also simulated the impact on the Rossiter-McLaughlin effect by including two star spots in the model, each with different locations and sizes (yielding a combined $\sim5$\,\% variation in flux over a stellar rotation). We find that the impact on the Rossiter-Mclaughlin signal from the two spot models to be even smaller than the single star spot models.

Given the relatively high uncertainty on each of the radial velocity measurements obtained with {\textsc{Minerva}}-Australis ($\gtrsim20$\,\mos), we do not expect star spots to have a significant influence on the recovery of $\lambda$ from the Rossiter-McLaughlin effect unless AU Mic\,b happens to cross over a star spot during one of the observed transits.

\begin{deluxetable*}{cccc}
\tabletypesize{\small}
\tablecaption{System Parameters, Priors, and Results for AU\,Mic\label{table:results}}
\tablehead{
\colhead{Parameter} & \colhead{Prior}  & \colhead{Results (linear model)} & \colhead{Results (polynomial model)} \\
\colhead{} & \colhead{}  & \colhead{} & \colhead{\textbf{preferred solution}}}
\startdata
  Planet-to-star radius ratio, $R_{P}/R_{\star}$ & $\mathcal{N}(0.0514;0.0013)$\tablenotemark{a} & $0.0513\pm0.0012$ & $0.0513\pm0.0012$ \\
  Mid-transit epoch (2450000-BJD), $T_{0}$ & $\mathcal{N}(8330.39153;0.00070)$\tablenotemark{a}  & $8330.39151\pm0.00064$ & $8330.39149\pm0.00064$ \\
  Orbital period, $P$ (days) & $\mathcal{N}(8.46321;0.00004)$\tablenotemark{a} & $8.46321\pm0.00004$ & $8.46321\pm0.00004$ \\
  Inclination angle, $I$ (deg)  & $\mathcal{N}(89.5;0.5)$\tablenotemark{a} & $89.6^{+0.4}_{-0.5}$ & $89.3^{+0.5}_{-0.4}$ \\
  Semimajor axis to star radius ratio, $a/R_{\star}$ & $\mathcal{N}(19.1;1.8)$\tablenotemark{a} & $19.7^{+1.6}_{-1.5}$ & $19.8^{+1.5}_{-1.4}$ \\
  Limb-darkening coefficient, $q_{1}$ & $\mathcal{U}(0.47;0.0;1.0)$\tablenotemark{b} & $0.53^{+0.30}_{-0.32}$ & $0.57^{+0.28}_{-0.34}$ \\
  Limb-darkening coefficient, $q_{2}$ & $\mathcal{U}(0.40;0.0;1.0)$\tablenotemark{b} & $0.49^{+0.32}_{-0.31}$ & $0.54^{+0.30}_{-0.33}$ \\
  RV semi-amplitude, $K$ (\mos) & 0\tablenotemark{c} & ... & ... \\
  Orbital eccentricity, $e$  & 0\tablenotemark{a} & ... & ... \\
  Argument of periastron, $\omega$ (deg)  & ... & ... & ... \\
  RV jitter 1st transit, $jit_1$ (ln \mos) & $\mathcal{U}(3.5;0.1;6.9)$ & $3.6\pm0.2$ & $3.5\pm0.2$ \\
  RV jitter 2nd transit, $jit_2$ (ln \mos) & $\mathcal{U}(3.1;0.1;6.9)$ & $2.9\pm0.2$ & $2.9^{+0.2}_{-0.3}$ \\
  RV jitter 3rd transit, $jit_3$ (ln \mos) & $\mathcal{U}(3.0;0.1;6.9)$ & $1.2^{+0.8}_{-0.7}$ & $1.1^{+0.8}_{-0.7}$ \\
  Stellar rotation velocity, $v\sin i_{\star}$ (\kms) & $\mathcal{N}(12.2;3.5)$\tablenotemark{d} & $10.6^{+2.1}_{-2.0}$ & $11.3^{+3.1}_{-2.7}$ \\
  Projected spin-orbit angle, $\lambda$ (deg)  & $\mathcal{U}(0;-180;180)$ & $-2^{+27}_{-26}$ & $47^{+26}_{-54}$
\enddata
\tablecomments{$\mathcal{N}(\mu;\sigma)$ is a normal distribution with mean $\mu$ and width $\sigma$, $\mathcal{U}(s;a;b)$ is a uniform prior with a starting value $s$ and lower and upper limits of $a$ and $b$, respectively.}
\tablenotetext{a}{Priors from Plavchan et al. (2020).}
\tablenotetext{b}{Initial values from interpolated values from look-up tables in Claret \& Bloemen (2011). }
\tablenotetext{c}{Radial velocity semi-amplitude is incorporated into the hybrid linear and polynomial fitting and is therefore fixed to zero.}
\tablenotetext{d}{An inflated 5\,$\sigma$ prior was applied on $v\sin i_{\star}$ as measured from the {\textsc{Minerva}}-Australis spectra to allow the MCMC to properly sample the posterior distribution.}
\end{deluxetable*}

\section{Discussion \& Conclusions} \label{sec:Discussion}
We have detected a marginal Rossiter–McLaughlin effect signal of AU\,Mic\,b from radial velocity transit observations using the {\textsc{Minerva}}-Australis telescope array. From these observations, we measured the sky-projected spin-orbit angle between the planet's orbit and the host star's spin axis. Due to AU Mic's extreme youth ($\sim23$\,Myr) and significant stellar activity compounded with our observations lacking full transit coverage (for two out of the three transits observed) plus sufficient out-of-transit data, we are unable to precisely measure the planet's orbital obliquity. We find $\lambda = 47{^{+26}_{-54}}^{\circ}$ for the preferred (and more conservative) hybrid second order polynomial activity model and $\lambda = -2{^{+27}_{-26}}^{\circ}$ for the hybrid linear activity model.

AU\,Mic is the youngest exoplanetary system for which a measurement of the spin-orbit alignment has been attempted, and one of only two systems younger than 100\,Myr with such measurements as of June 2021 \citep[the other being DS Tuc A, see,][]{2019ApJ...880L..17N,2019A&A...630A..81B,2020ApJ...892L..21Z,2020AJ....159..112M}. The vast majority of obliquity measurements to date have been made for hot Jupiters orbiting earlier type and older main-sequence stars \citep[e.g., ][]{2015ARA&A..53..409W}. Therefore, AU\,Mic occupies a unique parameter space and is an excellent laboratory for testing models of planet formation and misalignment.


Given the potential low-obliquity measured for AU\,Mic\,b in this study as well as confirmed to be of low-obliquity by other studies submitted in parallel with this one, it is likely that the planet formed beyond the ``ice-line'' within the protoplanetary disk around AU\,Mic and then migrated inwards as a result of its interaction with that disk \citep[e.g., see,][]{Disk1,Ice2,Disk4,cooljupiters} to its current $P\sim8.5$\,day orbit \citep[for an opposing view on the in situ formation of Jovian planets, see e.g.,][]{2016ApJ...829..114B}. Further evidence to support this comes from the fact that AU\,Mic's debris disk is observed to be nearly edge-on from far-infrared and sub-millimeter direct imaging \citep[e.g., see][]{2015ApJ...811..100M}, with a small aspect ratio suggesting a dynamically cold planetesimal population \citep{2015A&A...581A..97S,2019ApJ...875...87D}. Since AU\,Mic\,b transits its host star, this strongly suggests that the planet and the disk lie in the same orbital plane. This then increases the likelihood that the stellar inclination is also close to $90^{\circ}$ (i.e., the stellar equator is edge-on), since it can be expected that protoplanetary disks from which planets form (and the debris disks that mark the remnants of those disks at later epochs) should be orthogonal to the stellar angular momentum vector \citep[though see,][for a mechanism on perturbing a protoplanetary disk out of alignment at the epoch of star and planet formation]{2015ApJ...800..138N} as a consequence of the stellar formation process \citep{1964ApJ...139.1217T,Ice1}. Such star-planet-debris disk alignments have been observed for other planetary systems, such as HD 82943 \citep{2013MNRAS.436..898K}. Therefore, the sky-projected spin-orbit angle likely represents the true orbital obliquity of the system, and the debris disk, planetary orbit, and stellar equator all seem to be well-aligned.


It therefore appears unlikely that this planet experienced high-eccentricity driven migration in the past (e.g., planet-planet scattering, \citealt{2008ApJ...686..621F}, or Lidov-Kozai cycling with tidal friction, \citealt{2007ApJ...669.1298F}) given the low orbital obliquity and its youth, but instead sedately migrated inwards via disk-migration mechanisms \citep{Disk1}. Giant planet formation by the core-accretion model together with type 1 and 2 disk migration to short period ($<~10$\,day) orbits are predicted to operate on timescales of less than 10\,Myr \citep[see, e.g.,][]{2003ApJ...598L..55R,2005SSRv..116...53W,2013apf..book.....A}. Since AU\,Mic is a member of the $\beta$ Pictoris moving group, the star's age is well constrained at 23~$\pm$~3~Myr \citep{2014MNRAS.445.2169M}. Therefore, the planet's formation by core-accretion and subsequent migration via type 1 and 2 disk migration are completely compatible with the observations.

AU\,Mic now joins the ranks of the few systems that are known to host both planetary and planetesimal components \citep[e.g.][]{2018MNRAS.476.4584K,2020Yelverton}, making it an even more important analogue to the Solar System for studying the interplay between planetary and debris components. Furthermore, determining the obliquity distribution of young planetary systems like AU\,Mic will be crucial in establishing their formation and migration histories, dynamical processes which have a substantial impact on their architectures.

The all-sky transiting exoplanet survey \textit{TESS} has begun delivering new discoveries of young exoplanets orbiting bright stars that are needed to establish this obliquity distribution, and in the years to come, it is likely that systems such as AU\,Mic will prove pivotal in placing the formation of our own planetary system in context.

\acknowledgments
We respectfully acknowledge the traditional custodians of all lands throughout Australia, and recognise their continued cultural and spiritual connection to the land, waterways, cosmos, and community. We pay our deepest respects to all Elders, ancestors and descendants of the Giabal, Jarowair, and Kambuwal nations, upon whose lands the {\textsc{Minerva}}-Australis facility at Mt Kent is situated.

{\textsc{Minerva}}-Australis is supported by Australian Research Council LIEF Grant LE160100001, Discovery Grant DP180100972, Mount Cuba Astronomical Foundation, and institutional partners University of Southern Queensland, UNSW Australia, MIT, Nanjing University, George Mason University, University of Louisville, University of California Riverside, University of Florida, and The University of Texas at Austin.

JPM acknowledges research support by the Ministry of Science and Technology of Taiwan under grants MOST107-2119-M-001-031-MY3, MOST107-2119-M-001-031-MY3, and MOST109-2112-M-001-036-MY3, and Academia Sinica under grant AS-IA-106-M03.
MNG acknowledges support from MIT's Kavli Institute as a Juan Carlos Torres Fellow.

%

\vspace{5mm}
\facilities{{\textsc{Minerva}}-Australis}









\bibliography{references}{}
\bibliographystyle{aasjournal}


\clearpage
\newpage

\appendix

\begin{figure*}
  \includegraphics[width=\linewidth]{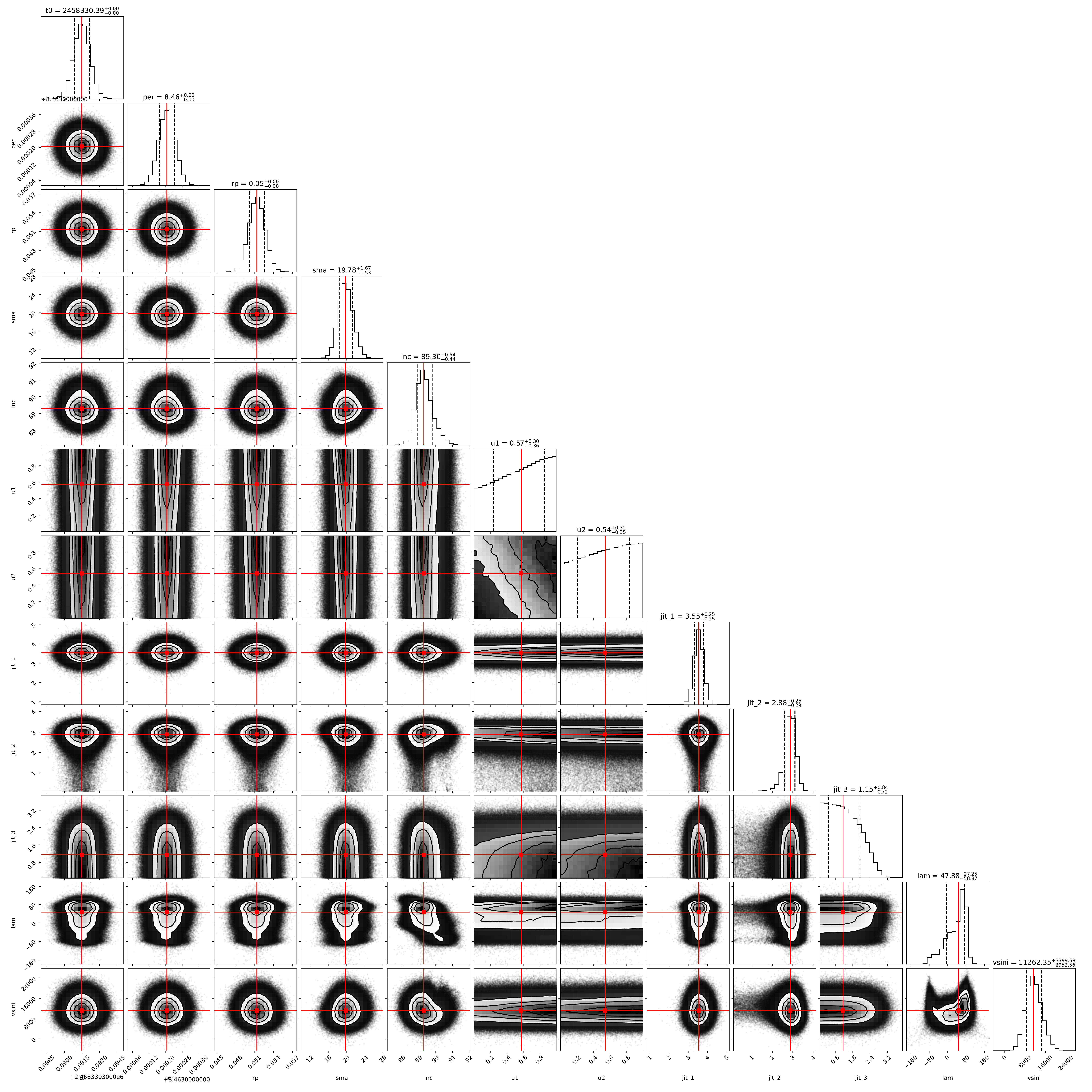}
  \caption{Corner plot of the posteriors using the second order hybrid polynomial activity model.}
  \label{fig:posterior}
\end{figure*}

\begin{figure*}
  \includegraphics[width=\linewidth]{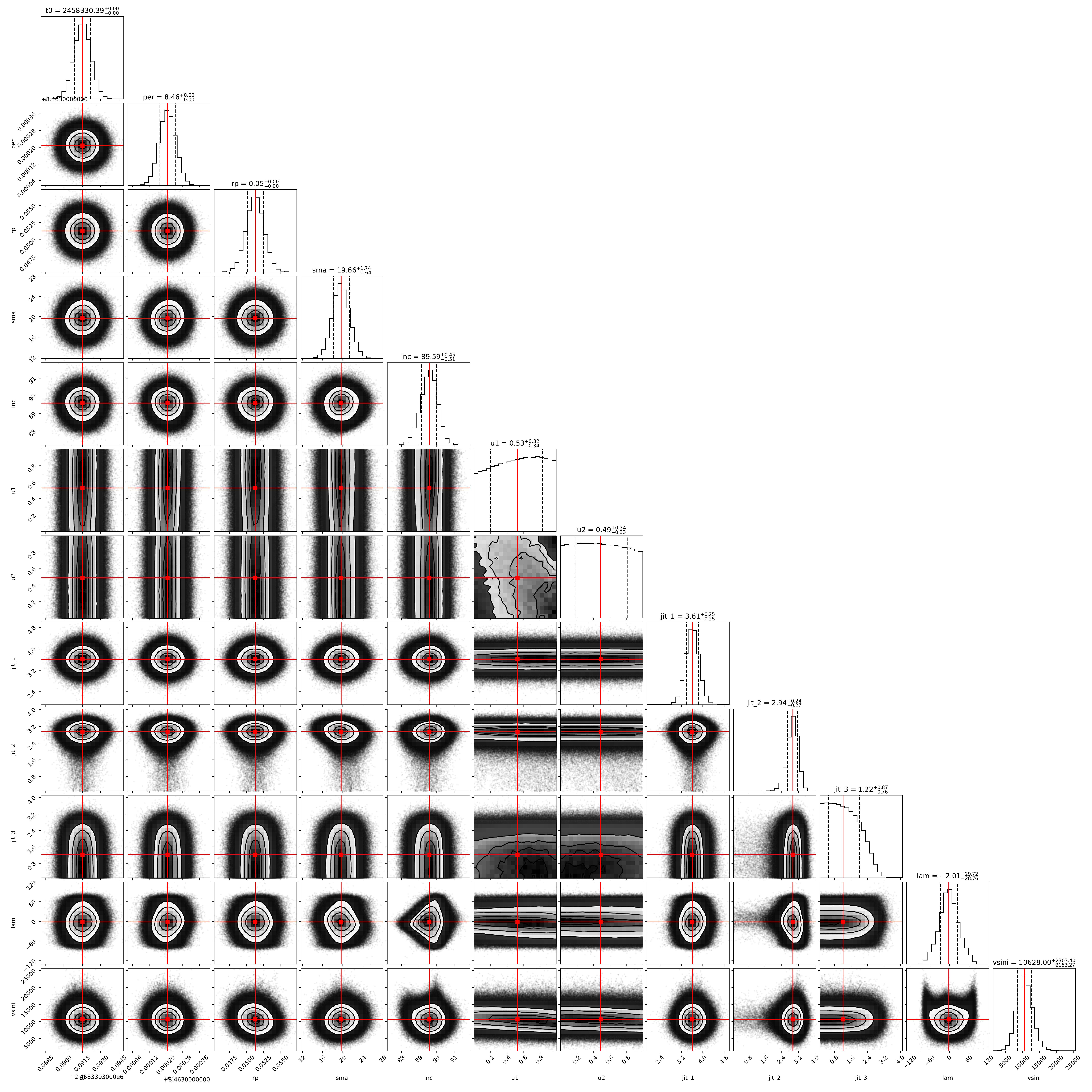}
  \caption{Corner plot of the posteriors using the hybrid linear activity model.}
  \label{fig:posterior_linear}
\end{figure*}



\end{document}